# Routing Dynamics in Distributed Quantum Networks

Mst Shapna Akter[1], Md. Shazzad Hossain Shaon[1], Tasmin Karim[1], Md. Fahim Sultan[1], and Emran Kanaan[1]

[1]Department of Computer Science and Engineering, Oakland University, Rochester, MI 48309, USA ,
Email: akter@oakland.edu, shaon@oakland.edu, tasminkarim@oakland.edu, mdfahimsultan@oakland.edu, emrankanaan@oakland.edu

## Abstract

Distributed quantum networks are not merely information conduits but intricate systems that embody the principles of quantum mechanics. In our study, we examine the underlying mechanisms of quantum connectivity within a distributed framework by exploring phenomena such as superposition and entanglement and their influence on information propagation. We investigate how these fundamental quantum effects interact with routing strategies that, while inspired by classical methods, must contend with quantum decoherence and measurement uncertainties. By simulating distributed networks of 10, 20, 50 and 100 nodes, we assess the performance of routing mechanisms through metrics that reflect both quantum fidelity and operational efficiency. Our findings reveal that the quantum coherence inherent in entangled states can enhance routing fidelity under specific conditions, yet also introduce challenges such as increased computational overhead and sensitivity to network scale. This work bridges the gap between the underlying principles of quantum systems and practical routing implementations, offering new insights into the design of robust distributed quantum networks.

## 1 Introduction

Distributed quantum networks represent a revolutionary paradigm in the evolution of communication systems, merging the counterintuitive tenets of quantum mechanics with advanced network science. In these systems, nodes distributed over a network are interconnected not merely by classical signals but by quantum states that exhibit superposition and entanglement [1]. These quantum phenomena allow qubits to exist in multiple states simultaneously and maintain non-classical correlations across distant nodes, thus offering unprecedented opportunities for secure communication and parallel information processing [2]. However, the promise of such networks is accompanied by formidable challenges, primarily due to the inherent fragility of quantum states in the presence of environmental disturbances. In particular, decoherence—the loss of quantum coherence resulting from interactions with the environment—leads to the degradation of entangled states and reduces the fidelity of transmitted information [3]. Furthermore, measurement uncertainties intrinsic to quantum systems add another layer of complexity, making the task of reliably routing quantum information significantly more challenging compared to classical networks.

Routing in distributed quantum networks must, therefore, address not only the traditional demands of efficiency and resource optimization but also the preservation of delicate quantum correlations. Our study systematically investigates this intricate balance by evaluating a set of routing protocols under realistic network conditions. We construct scalable models of networks with 10, 20, 50, and 100 nodes to analyze how quantum phenomena interact with routing strategies. Specifically, we examine four distinct algorithms: a genetic algorithm inspired by evolutionary dynamics, two deterministic methods derived from classical shortest-path paradigms, and a reinforcement learning approach that adapts dynamically to network conditions. Each method is evaluated using metrics that capture both quantum fidelity—reflecting the integrity of the entangled states—and operational efficiency in terms of execution time and path length.

The key contributions of this work can be summarized as follows:

- A. We develop an analytical framework that integrates stochastic models of entanglement generation with decoherence effects to derive closed-form expressions for link-level entanglement generation time and entanglement rate.

- B. We propose and adapt multiple routing protocols for distributed quantum networks, including deterministic and adaptive strategies, and incorporate these into a dynamic simulation framework.

- C. We conduct extensive numerical simulations across networks of varying sizes to systematically compare the performance of the proposed routing protocols in terms of quantum fidelity, execution time, and path length.

- D. Our findings provide insights into the trade-offs between computational efficiency and the preservation of quantum coherence, offering guidance for the design of robust and scalable quantum communication infrastructures.

The remainder of the paper is organized as follows. Section II, reviews recent advances and challenges in the field of quantum network routing. Section III, describes our network model along with the assumptions and definitions that underpin our analytical framework. Section IV, presents the derivation of performance metrics and offers a comprehensive numerical evaluation of the proposed routing protocols. Section V, discusses the constraints and limitations of our approach. Finally, Section VI, summarizes our key findings and outlines potential directions for future research in distributed quantum networks.

## 2 Related Works

Recent literature has provided significant improvements in recognizing and conquering the issues of entanglement in quantum networks. This study looks at the technical and conceptual changes required to scale up quantum communication networks from individual localized links to a globally associated quantum network structure [4]. In another study, the authors address the challenges and opportunities of routing in quantum networks. They categorize existing routing approaches into two categories: fundamental routing and networking that includes connection purifying [5]. Another study, particularly that addressed high-fidelity entanglement routing in quantum communications, introduced a novel Purification-enabled Entanglement Routing Algorithm (PERA). The initial solution of PERA uses network throughput as a routing metric and performs a hop-by-hop purification procedure to ensure integrity at each node along the route [6]. Another research aims to find strategies that reduce network latency while forming entanglement between two geographically dispersed nodes. The authors provided three new routing algorithms with analytical information to show that, for single requests, these algorithms achieve lower latency on a continuous entanglement distribution model than an on-demand approach[7]. Muralidharan et al. [8] describe the first comprehensive comparison of three generations of quantum repeaters. The authors assessed the temporal and physical resource costs associated with each generation before determining the best quantum repeater design for the quantum key distribution given specified experimental conditions. Wallnöfer et al.[9] development of quantum communication protocols is treated as a reinforcement learning (RL) challenge. RL and machine learning (ML) in ggeneralre proving more useful for automating problem solving in quantum information science. This work indicates to provide insight into the challenges and open concerns regarding Quantum Internet architecture. It starts by explaining the fundamental quantum physics ideas required to distinguish between classical and quantum networks. Quantum teleportation is then introduced as the primary method for delivering quantum information without physically moving the information carrier, in accordance with quantum mechanical principles. Finally, the primary research issues associated with developing quantum communication networks are discussed [10]. In another study, researchers efficiently and reliably teleported a controlled-Z (CZ) gate across two circuit qubits in separate modules, achieving 86% fidelity. Following that, they used Grover's search method [11]. Another study found that dispersing entanglement among network nodes improves scalability as the network size grows. A shared quantum non-demolition measurement was utilized to entangle a clock network with up to four nodes [12].

## 3 Materials and Methods

In our study, we model the distributed quantum network as an undirected graph [13]

$$G = (V, E)$$

where the set $V$ comprises quantum nodes and $E$ represents the optical links that interconnect these nodes. Each node is equipped with a quantum processor and quantum memories, while the links simulate optical fibers over which entangled photons are transmitted. Quantum repeaters [14] are integrated into the network architecture to enable entanglement swapping across adjacent links, thereby extending the reach of entanglement distribution without suffering from the exponential decay associated with long-distance transmission.

The process of entanglement generation at each node is modeled as a stochastic event [15]. Specifically, an atom–photon entanglement is generated with a probability $p$, and the operation requires an average time $\tau_p$. Once an entanglement is generated, it is transmitted along the optical link, and the success of this process is further conditioned on factors such as detector efficiencies and channel attenuation. To quantify the link-level performance, we derive an expression for the average entanglement generation time between two adjacent nodes $i$ and $j$. This time, denoted as $T_{(i,j)}$, is given by

$$T_{(i,j)} = \frac{1 - p_{(i,j)} \cdot T_{f_{(i,j)}} + p_{(i,j)} \cdot T_{s_{(i,j)}}}{p_{(i,j)}},$$

where $p_{(i,j)}$ is the probability of successful entanglement generation along link $(i, j)$, $T_{f_{(i,j)}}$ is the average time for a failed entanglement attempt (which includes the cooling time $\tau_d$ required to reset the node), and $T_{s_{(i,j)}}$ represents the time for a successful entanglement attempt. The entanglement rate $\xi_{(i,j)}(T_{ch})$ is then computed as the reciprocal of $T_{(i,j)}$, provided that the elapsed time since entanglement generation does not exceed the quantum memory coherence time $T_{ch}$.

Our simulation framework integrates this stochastic model of entanglement generation with a dynamic routing module. The framework computes the entanglement rate for each link and aggregates these rates along candidate paths to determine the end-to-end entanglement performance. Networks with varying numbers of nodes (10, 20, 50, and 100) are simulated to study the impact of network scale on both the routing efficiency and the preservation of quantum fidelity. In parallel, our framework incorporates classical communication delays, which are modeled as a function of the physical length of the optical links, thereby accounting for the time required for acknowledgment messages in entanglement swapping operations.

Routing protocols are implemented by adapting classical algorithms to the quantum domain. In particular, deterministic methods such as Dijkstra's and Bellman-Ford algorithms are modified to incorporate the computed entanglement rates as the routing metric. Additionally, adaptive protocols based on genetic algorithms and reinforcement learning are implemented to explore diverse routing paths and optimize the trade-off between computational overhead and the preservation of quantum coherence. For each routing protocol, the performance is evaluated using metrics such as quantum fidelity, entanglement rate, execution time, and the number of hops (path length). All the steps are shown in Fig 1.

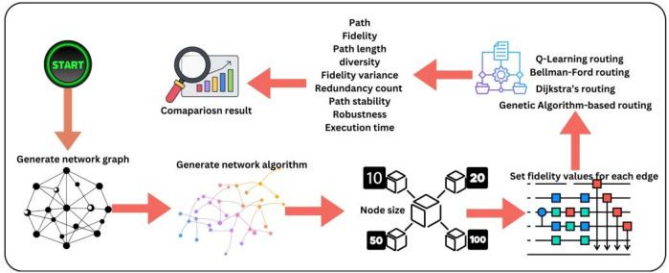

Figure 1: Overall working procedure of the study.

## 4 Results and Analysis

Our simulation framework was used to evaluate the performance of four routing protocols in distributed quantum networks of varying sizes (10, 20, 50, and 100 nodes). In the experiments, we compared a Genetic Algorithm (GA), Dijkstra's Algorithm, the Bellman-Ford Algorithm, and a Q-Learning based approach. The key performance metrics considered include the end-to-end quantum fidelity, the execution time (in seconds), and the number of hops (path length) of the selected route. All methodologies and assessments in this study were carried out with 64 GB RAM, a Intel Ultra 9 CPU, 32 GB VRAM, and a dedicated 12 GB NVIDIA RTX 4080 GPU. The implementation utilized Qiskit for quantum computing simulations and NetworkX (nx-graph) for quantum network modeling and performance evaluations.

For a network comprising 10 nodes, the GA produced a path [0, 1, 9] with a fidelity of 0.9655, an execution time of 0.021 seconds, and a path length of 3. In contrast, both Dijkstra's and Bellman-Ford algorithms yielded a shorter path [0, 9] with a slightly lower fidelity of 0.9335, zero execution time, and a path length of 2. The Q-Learning protocol, however, demonstrated erratic behavior by repeatedly selecting looping paths (e.g., [0, 1, 0, 1, ..., 9]) and resulted in a significantly lower fidelity of 0.4280. When the network size was increased to 20 nodes, the GA maintained a high fidelity of 0.9820 (via the path [0, 15, 19]) with a modest execution time (0.024 seconds) and a path length of 3. Both Dijkstra's and Bellman-Ford algorithms continued to produce a consistent path [0, 19] with fidelities of 0.9310 and zero execution time, while the Q-Learning approach's performance further deteriorated, showing a fidelity of only 0.1915. For networks with 50 nodes, all three deterministic methods (Dijkstra's, Bellman-Ford, and GA) converged to a similar performance in terms of fidelity (0.9094) and execution time (0.022 seconds for GA, and 0.0 seconds for the deterministic methods) with a path length of 2. In contrast, the Q-Learning protocol, though somewhat improved relative to the 20-node case, achieved a fidelity of only 0.6299, reflecting persistent instability. The trend becomes even more pronounced in 100-node networks. Here, the GA achieved a fidelity of 0.9264 along the path [0, 20, 27, 44, 5, 63, 99] with an execution time of 0.046 seconds and a path length of 7. Notably, both Dijkstra's and Bellman-Ford algorithms exhibited high fidelities of 0.9779 with very short execution times (0.003 and 0.004 seconds, respectively) and a minimal path length of 2. In stark contrast, the Q-Learning approach deteriorated dramatically, with its fidelity dropping to 0.0164, indicating severe performance degradation in large-scale networks.

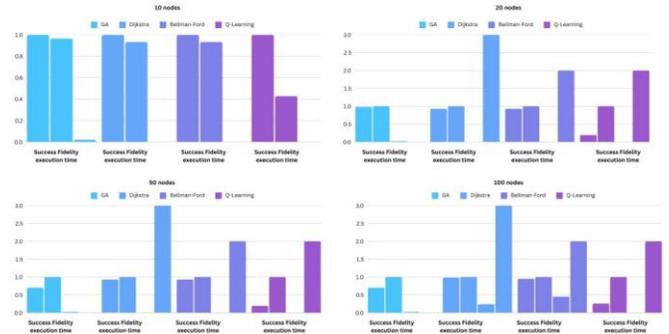

Figure 2: Four different algorithms result based on 10, 20, 50, and 100 nodes

The results indicate that deterministic protocols, namely Dijkstra's and Bellman-Ford algorithms, consistently yield the shortest paths and minimal execution times, which is advantageous for rapidly establishing routing decisions demonstrates in Fig 2. However, these protocols sometimes sacrifice quantum fidelity, especially in smaller networks, where the GA is observed to outperform them by providing higher fidelity through more diverse path selections. Notably, as the network size increases, the GA continues to deliver competitive fidelity, although at the cost of longer paths and slightly increased execution times. In contrast, the Q-Learning approach, despite its theoretical adaptability, suffers from excessive looping and instability, leading to drastic performance degradation, particularly in large-scale (100-node) networks.

## Discussion

Our experimental evaluation of routing protocols in distributed quantum networks has revealed several important insights into the interplay between quantum fidelity and routing efficiency. Deterministic protocols such as Dijkstra's and Bellman-Ford algorithms, while computationally efficient and capable of rapidly identifying the shortest paths, exhibit limitations in preserving the quantum state fidelity necessary for high-quality entanglement distribution. This trade-off becomes evident when comparing their performance to that of the Genetic Algorithm (GA), which, despite incurring a modest increase in both path length and execution time, consistently delivers higher fidelity across networks of various sizes. The GA's ability to explore diverse routing paths appears to mitigate some of the adverse effects of decoherence by selecting routes that, while not always the shortest in terms of hops, maintain stronger quantum correlations over the distributed network. In contrast, the Q-Learning approach demonstrated significant instability, particularly in larger networks, where the algorithm's propensity for looping led to a marked degradation in performance. This instability underscores the challenge of adapting reinforcement learning techniques to the inherently stochastic and noise-sensitive environment of quantum networks. Moreover, our results highlight the critical influence of network scale on routing performance. As the number of nodes increases, the complexity of maintaining high quantum fidelity grows, necessitating more sophisticated routing strategies that can dynamically balance the conflicting demands of efficiency and quantum coherence preservation. The observed performance degradation of Q-Learning in 100-node networks, in particular, suggests that further refinement—potentially through hybrid strategies that integrate deterministic and adaptive approaches—may be required to achieve robust routing in large-scale quantum systems. Our analysis demonstrates that while deterministic routing protocols offer simplicity and speed, they may not always suffice when quantum fidelity is the paramount concern. Adaptive strategies, as exemplified by the GA, show promise in addressing the unique challenges of distributed quantum networks, though they come with increased computational overhead. Future work will focus on enhancing these adaptive methods, exploring hybrid routing protocols, and further investigating the impact of network topology and dynamic environmental factors on the performance of quantum routing strategies.

Table 1: Performance Comparison for Different Network Sizes

| Nodes | Algorithm | Path | Fidelity | Exec. Time (s) |
|---|---|---|---|---|
| 10 | GA | [0, 1, 9] | 0.9655 | 0.021 |
| | Dijkstra | [0, 9] | 0.9335 | 0.0 |
| | Bellman-Ford | [0, 9] | 0.9335 | 0.0 |
| | Q-Learning | [0, 1, 0, 1, ..., 9] | 0.4280 | – |
| 20 | GA | [0, 15, 19] | 0.9820 | 0.024 |
| | Dijkstra | [0, 19] | 0.9310 | 0.0 |
| | Bellman-Ford | [0, 19] | 0.9310 | 0.0 |
| | Q-Learning | [0, 15, 0, 15, ..., 19] | 0.1915 | – |
| 50 | GA | [0, 49] | 0.9094 | 0.022 |
| | Dijkstra | [0, 49] | 0.9094 | 0.0 |
| | Bellman-Ford | [0, 49] | 0.9094 | 0.0 |
| | Q-Learning | [0, 32, 0, 32, ..., 49] | 0.6299 | – |
| 100 | GA | [0, 20, 27, 44, 5, 63, 99] | 0.9264 | 0.046 |
| | Dijkstra | [0, 99] | 0.9779 | 0.003 |
| | Bellman-Ford | [0, 99] | 0.9779 | 0.004 |
| | Q-Learning | [0, 50, 0, 50, ..., 99] | 0.0164 | – |

## 5 Limitation

This is based on a quantum-based simulation called QISKIT, and we utilized certain library packages for the analysis, as well as an Anaconda Jupiter virtual environment on a 64GB RAM machine. Analyzing quantum networking limits on resources for long-distance communication on an average computer requires simulation and modeling, as compared with actual establishing a quantum network. Quantum networks require a enough number of high-quality qubits at each node. Currently, the number of useful qubits in quantum devices remains restricted. Quantum activities, such as entanglement creation and manipulation, are not flawless. Errors occur, and the accuracy of these procedures is critical for ensuring trustworthy quantum communication. Entanglement swapping, or extending entanglement across larger distances, necessitates high-fidelity quantum computations and efficient quantum memory. The quality of stored qubits can decline with time, causing mistakes in quantum communication protocols. Storing the state vector of a quantum system necessitates a vast quantity of classical memory. For example, simulating 50 qubits takes terabytes of RAM. Quantum simulators are useful tools, but they have certain inherent limits. Qiskit's simulators, such as Aer, have the inherent constraints of classical quantum system simulations. They struggle to handle huge qubit quantities. Qiskit's transpiler tries to optimize circuits for various backends, however it is not always successful. Simulating quantum networks, particularly ones with complicated topologies and protocols, is computationally demanding. NetSquid, while efficient, nevertheless has scaling issues. NetSquid provides models, but their correctness is contingent on the underlying assumptions and parameters. NetSquid provides models, but their correctness is contingent on the underlying assumptions and parameters. While Cirq supports noise models, creating accurate representations of real-world noise remains a challenge. The majority of research institutions do not have access to IBQM computers, which makes it difficult for researchers to properly locate the network using a software simulator.

# 6 Conclusion

In this study, we developed a comprehensive simulation framework to evaluate routing protocols in distributed quantum networks. By modeling the network as an undirected graph where nodes represent quantum processors and optical links serve as channels for entangled photon transmission, we captured the inherent stochastic nature of entanglement generation and the challenges imposed by decoherence. Our analytical model provided closed-form expressions for link-level entanglement generation time and entanglement rate, which were subsequently incorporated into our routing evaluation. The simulation results demonstrate that deterministic protocols, such as Dijkstra's and Bellman-Ford algorithms, consistently yield short paths and minimal execution times. However, these methods tend to provide lower quantum fidelity compared to adaptive techniques. The Genetic Algorithm (GA) emerges as a strong candidate by balancing high fidelity and routing diversity, even though it incurs a moderate increase in path length and execution time. In contrast, the Q-Learning approach, while theoretically promising, exhibits excessive looping and instability, particularly in larger networks, leading to significantly degraded performance. Overall, our findings highlight the critical trade-offs between computational efficiency and the preservation of quantum coherence in routing for distributed quantum networks. The insights derived from this work provide a foundation for designing robust routing protocols that can effectively support scalable quantum communication infrastructures. Future work will focus on refining adaptive strategies and exploring hybrid protocols that combine the strengths of both deterministic and evolutionary approaches to further enhance routing performance in complex quantum networks.